\documentstyle[multicol,prl,aps]{revtex}
\begin{document}

%%%%%%%%%%%%%%%%%%%%%%%%%%%%%%%%%%%%%%%%%%%%%%%%
\title{ 
Comment on ``On the Renormalization of Conductance
in Tomonaga-Luttinger Liquid''
}

\author{Satoshi Fujimoto$^{1}$ and Norio Kawakami$^{2,3}$
}
%%%%%%%%%%%%%%%%%%%%%%%%%%%%%%%%%%%%%%
\address
{$^{1}$Department of Physics, Kyoto University, Kyoto 606 \\
$^{2}$Department of Applied Physics, Osaka University, 
Suita, Osaka 565 \\
$^{3}$Department of Material and Life Science, 
Osaka University, Suita, Osaka 565
}
%\date{}
\maketitle

%%%%%%%%%%%%%%%%%%%%%%%%%%%%%%%%%%%%%%%
\begin{abstract}
Recently Kawabata found that the conductance 
in Tomonaga-Luttinger (TL) liquids should not be renormalized
if one incorporates the renormalization of 
applied fields correctly. This claim is generalized to 
include the interactions which 
carry the large momentum transfer. We thus
find that Kawabata's conclusion holds generally for the
class of TL liquids.
A key observation is that such 
irrelevant interactions merely renormalize the velocity and 
the TL parameter at the massless TL fixed point.
\end{abstract}
%%%%%%%%%%%%%%%%%%%%
\begin{multicols}{2}
%\newpage
%%%%%%%%%%%%%%%%%%%%%
In the recent paper, Kawabata has shown that the 
conductance in Tomonaga-Luttinger (TL) liquids
is not renormalized,  i.e.
$G=2e^2/h$, if one correctly 
takes into account the renormalization of 
applied fields self-consistently\cite{kawabata}.  This is 
a plausible scenario  for the observed conductance
\cite{exp}, which is distinct from the previous 
proposals\cite{stone}. 
He has demonstrated this fact by using a special version
of the TL model. He has also raised a question how his conclusion 
would be modified if 
the interactions carrying the large momentum transfer
are incorporated, but  this problem remains still open.
In this comment, we show that
Kawabata's results for the conductance hold even if 
such interactions are incorporated, 
and his conclusion is quite general for the 
TL liquids.

Let us start with the model 
studied by Kawabata \cite{kawabata},
%%%%%%%%%%%%%%%%%%%%%%%%%%%%%
\begin{eqnarray}
%%{\cal H}_0 &=& v_F \sum_{\alpha=\pm} 
%%\int dx \, : \psi^{\dag}_{\alpha}
%%(i \alpha{d \over dx}) \psi_{\alpha}(x):,
{\cal H}_0 &=&  {- \hbar^2 \over 2m } 
\int dx \, \psi^{\dag}(x)
{d^2 \over dx^2} \psi(x) \, , 
\nonumber \\
%%%%%%%%%%%%%%%%%%%%%%%%%%%%%
{\cal H}_1 &=& {g \over 2 \pi} \int dx \, :\rho(x)\rho(x): \, ,
\label{ham1}
\end{eqnarray}
%%%%%%%%%%%%%%%%%%%
where $\psi(x)$ is the Fermi field, 
$::$ denotes the normal ordering,
and  $\rho(x)=\rho_+(x)+ \rho_-(x)$ with 
 $\rho_+(x)$ ($\rho_-(x)$) being the density
of left- (right-) going electrons. 
This is a special version of the TL
model. For simplicity, we have omitted the
spin index since only the charge sector is relevant
to the conductance.  
In this translationally invariant model,
the renormalization of the Fermi velocity $v_F$ is
uniquely related to  the TL parameter $\gamma$ 
via $v_F \rightarrow v_F/\gamma$.
This makes the analysis very simple, and 
the self-consistent treatment\cite{kawabata}
turns out to be exact.

If we take into account more general interactions,
the velocity and the TL parameter are renormalized 
in different manners. We consider the general model describing
TL liquids for which the interaction term is given by,
%%%%%%%%%%%%%%%%%%%%%%%%%%%%%%%
\begin{eqnarray}
{\cal H}_1 = {1 \over 2\pi} &  \int & dx[g_4 \sum_{\alpha=\pm}
: \rho_\alpha(x)\rho_\alpha(x): \nonumber \\
& + & 2 g_2 : \rho_+(x)\rho_-(x):] + {\cal H}_U,
\label{ham2}
\end{eqnarray}
%%%%%%%%%%%%%%%%%%%
where ${\cal H}_U$ denotes  the interactions other than
forward scatterings, which may 
include the large momentum transfer. 
%%%%%%%%%%%%%%%%%%%%%%%%%
Though it seems difficult at first glance to 
study the effect of the interactions (\ref{ham2})
on the conductance, this problem can be
resolved in the following way.  
%%%%%%%%%%%%%%%%%%%%%%%%%%
The first key observation is that ${\cal H}_U$ should become 
irrelevant after the renormalization 
so far as the system remains as TL liquids.
Hence, what is caused by ${\cal H}_U$ is just   
to renormalize the velocity $ v_F \rightarrow v_F^*$
and the interactions $g_2(g_4)\rightarrow g_2^*(g_4^*)$
in different ways. 
This is a characteristic feature which does not 
appear in the special model (\ref{ham1}) \cite{kawabata}.
%%The renormalized  Hamiltonian 
%%is exactly solvable at the TL fixed point.
The renormalized  Hamiltonian still has two different
types of the interactions $g_2^*$ and $g_4^*$.
For our purpose, it is crucial
to  rewrite the renormalized 
Hamiltonian in a suitable form to apply the analysis
of Kawabata. Namely, passing to the 
continuum limit, the renormalized Hamiltonian
including the kinetic-energy term 
is shown to be equivalent to 
%%%%%%%%%%%%%%%%%%%%%%%%%%%%%%%%%%
\begin{eqnarray}
\widetilde { \cal H} = \hbar \widetilde v_F
\sum_{\alpha=\pm} && \int dx \,  : \psi^{\dag}_{\alpha} 
(i \alpha{d \over dx}) \psi_{\alpha}(x): \nonumber\\
  && + {\widetilde g \over 2 \pi}
 \int dx \, :\rho(x)\rho(x): \, ,
\label{ham3}
 \end{eqnarray}
%%%%%%%%%%%%%%%%%%%%%%%%%%%%%%%%%%%%%%%%%%%%%
where  $ \psi^{\dag}_{\pm}$ is the right- (left-)
going Fermi fields, $\hbar \widetilde v_F=
\hbar v_F^* + (g_4^*-g_2^*)/\pi$
and $\widetilde g= g_2^*$. Here we have used the fact
that the $g_4^*$ term merely renormalizes the velocity.
One can now see that the reduced Hamiltonian (\ref{ham3})
has the effective interaction which is translationally 
invariant, and is essentially the same as (\ref{ham1}) studied
by Kawabata, so that we can directly follow his arguments 
on the renormalization of applied fields \cite{com}.
Exploiting his results, we thus end up with the conclusion that the 
conductance in TL liquids is not renormalized, i.e. 
%%%%%%%%%%%%%%%%%%%%
\begin{equation}
G=2e^2/h,  
\end{equation}
%%%%%%%%%%%%%%%%%%%%
even if we take into account the interactions (\ref{ham2})
which may carry the large momentum transfer. Therefore Kawabata's
conclusion holds not only for the special model (\ref{ham1}) but
also for the class of TL liquids generally.

\end{multicols}

\begin{references}
\bibitem{com} The TL parameter is 
given  by  $\gamma=[1 + \widetilde g/(\pi \hbar 
\widetilde v_F)]^{-1/2}$.
\end{references}
\end{document}